# Microscopic origin of bipolar resistive switching of nanoscale titanium oxide thin films

Hu Young Jeong and Jeong Yong Lee [a]

*Department of Materials Science and Engineering, KAIST, Daejeon 305-701, Korea*

Sung-Yool Choi [b]

*Electronics and Telecommunications Research Institute (ETRI), Daejeon, 305-700, Korea*

Jeong Won Kim

*Korea Research Institute of Standards and Science (KRISS), Daejeon 305-340, Korea*

We report a direct observation of the microscopic origin of the bipolar resistive switching behavior in nanoscale titanium oxide films. Through a high-resolution transmission electron microscopy, an analytical TEM technique using energy-filtering transmission electron microscopy and an *in situ* x-ray photoelectron spectroscopy, we demonstrated that the oxygen ions piled up at top interface by an oxidation-reduction reaction between the titanium oxide layer and the top Al metal electrode. We also found that the drift of oxygen ions during the on/off switching induced the bipolar resistive switching in the titanium oxide thin films.

---
[a] Electronic mail: j.y.lee@kaist.ac.kr
[b] Electronic mail: sychoi@etri.re.kr

The bistable switching phenomenon, reversible switching between a high resistance state and a low resistance state, have been discovered for various binary oxide films such as NiO,[1] $Nb_2O_5$,[2] $Al_2O_3$,[3] $SiO_2$,[4] and $TiO_2$.[5] Since the recent publication by Baek et al.[6] in 2004, resistive random access memory (RRAM) using simple binary transition metal oxides such as NiO and $TiO_2$ has attracted extensive attention as a high-potential next-generation nonvolatile memory (NVM) due to its simple process, simple device structure, and high CMOS compatibility.[7-9] In the fundamental point of view, the electrical properties of nanoscale binary oxide thin films provides gives deeper understanding of the recently rediscovered memristors, the missing component of the basic circuits elements. Hewlett-Packard (HP) group[10,11] suggested that the motion of dopants or impurities, such as oxygen vacancies acting as mobile +2 charged dopants, was able to produce remarkable changes in the device resistance (memristic behavior), especially in $TiO_{2-x}$ devices. To verify the above models, it is crucial to identify the origin and movement of oxygen vacancies in the actual device structures. However, researchers have not yet fully shown a direct evidence of the movement of oxygen vacancies.

In this letter, we report on the presence and movement of oxygen vacancies formed by a redox reaction at the $Al/TiO_2$ top interface through a high-resolution transmission electron microscopy (HRTEM), an analytical TEM technique using energy-filtering transmission electron microscopy (EFTEM) and an in-situ x-ray photoelectron spectroscopy (XPS).

To fabricate the $Al/TiO_2/Al$ memory device, a $TiO_2$ oxide film with a thickness of ~16 nm was deposited on an $Al/SiO_2/Si$ substrate by the PEALD (ASM Genitech MP-1000) method at a substrate temperature of 180℃. The 50nm thick aluminum bottom and top electrodes were deposited by thermal evaporation method, forming the cross-bar type structures using a metal shadow mask with a line width of 60 μm. The cross-section images and chemical analysis of Al/TiO$_2$/Al samples were examined by HRTEM and analytical TEM. A 300 kV JEOL JEM 3010 with a 0.17 nm point resolution and a high-voltage electron microscope JEOL ARM-1300 operating at 1250 keV equipped with energy-filtering TEM were used. The XPS spectra were measured using Mg Kα radiation (ℏω = 1253.6 eV) and a SES-150 analyzer (Gamadata) equipped with a 2D CCD detector, both of which give a total energy resolution of 0.9 eV for each core level measurement. Using a thermal evaporation cell, the Al was deposited on TiO$_2$ film in a high vacuum chamber with a deposition rate of 0.25 nm min$^{-1}$. After the Al deposition, the sample was transferred to an analysis chamber without exposure to air for the XPS measurement. The electrical property (I-V curve) was measured using a Keithley 4200 Semiconductor Characterization System in a Dc sweep mode.

Figure 1 shows the typical J-V characteristic of an Al/TiO$_2$/Al memory cell measured at room temperature under the DC voltage sweep. The J-V curve exhibits a clear hysteretic and asymmetric behavior. Bistable resistance switching (BRS) between a high-resistance state (HRS) and a low-resistance state (LRS) was induced by the opposite polarity of the applied voltage. It was reported by Yu et al.[12] that this bipolar resistive switching (BRS) was associated with the charge trap sites in the top domain of the Ti oxide thin layer. The right bottom inset of Fig. 1 shows a cross-sectional bright-field (BF) TEM image of the Al/TiO$_2$/Al heterostructure with three distinctive layers (the Al bottom electrode, the Ti oxide thin layer, and the Al top electrode). It reveals that the highly uniform Ti oxide layer has been deposited due to an excellent roughness of an Al bottom electrode and the self-limited surface reactions of the atomic layer deposition method.

To clearly characterize the interface regions, we performed the HRTEM measurements of the sequential samples during the fabrication processes. Figure 2 is a series of cross-sectional HRTEM images of stacked Al/TiO$_2$/Al layers which were observed sequentially before and after depositing the titanium oxide thin layer. Note that the very thin native aluminum oxide with a 2~3 nm thickness was already been present before the ALD process, as shown in Figure 2(a). Figure 2(b) displays the HRTEM image of the surface of the Al bottom electrode annealed at 180℃, that is, the same temperature as in the ALD process chamber. It can be inferred that the native aluminum oxide expands to the ~3nm thickness due to a thermal annealing effect during the specimen loading process in the ALD main chamber. The uniform titanium oxide film was deposited on the native aluminum oxide with a 400 cycle ALD process. In Figure 2(c), it can be seen that the as-deposited titanium oxide layer with a dark contrast clearly has an amorphous phase with a thickness of ~16 nm. After the aluminum thermal evaporation process, the top interface layer was formed on the titanium oxide thin film, showing the amorphous phase with a light contrast, as shown in Figure 2(d). In addition, it is remarkable that the top domain of the titanium oxide layer also has a light contrast, as compared with Figure 2(c). (See the red dotted lines in Figures 2(c) and 2(d)). It is accepted that the composition of the upper Ti oxide layer has been drastically changed due to the reaction with the top aluminum metal layer during the deposition process. Generally, aluminum is well known as one of the most readily oxidizable metals. Therefore, aluminum atoms easily attract oxygen atoms from the titanium oxide layer and react with them, forming a new interfacial Al oxide layer.[13] This oxidation-reduction process induces the oxygen vacancies in the region of titanium oxide.

To confirm the chemical reaction between the amorphous titanium oxide and the aluminum top electrode, x-ray photoemission spectroscopy (XPS) was also estimated. Dake and Lad[14] reported that aluminum strongly reduced the single crystal TiO$_2$ surface and an aluminum oxide was formed at the interface using X-ray and ultra-violet photoelectron spectroscopies. However, it has yet not been confirmed whether a similar reaction could occur at metal-amorphous $TiO_2$ interfaces. For this experiment, an *in-situ* XPS measurement was carried out upon growing ultra-thin aluminum metal layers on the amorphous titanium oxide. Figure 3 shows the XPS spectra of Ti 2p and Al 2p measured as a function of aluminum deposition thickness. For the amorphous titanium surface, Ti cations are all in the $Ti^{4+}$ state at a binding energy of 458.9 eV, as shown in the lowest curve of Figure 3(a). By increasing the aluminum dose, a shoulder at the low binding energy side of Ti 2p spectra emerges and these spectra can be deconvoluted to show the contributions by a series of different oxidation states ($Ti^{3+}$, $Ti^{2+}$, $Ti^+$), as indicated in Figure 3(a). This clearly confirms that amorphous titanium oxide surface is reduced and the amount of reduced titanium oxide increases with additional aluminum doses. On the contrary, Figure 3(b) represents that the Al 2p spectra firstly appears in the $Al^{3+}$ state, which corresponds to the oxidized state of Al. The $Al^{3+}$ intensity increases until the aluminum thickness reaches 3 nm. However, a new Al metallic state ($Al^0$) appears at the thickness of 6.2 nm. This shows that aluminum oxide phases with a thickness of 3~6 nm are usually formed before metallic aluminum phases become stable, which is clearly consistent with the HRTEM images. The XPS results suggest that the aluminum is readily oxidized by extracting oxygen ions from the amorphous titanium oxide surface, causing titanium reduction.

The chemical change of an amorphous titanium oxide thin layer sandwiched between two aluminum electrode, in the real off and on states, was characterized using electron energy loss spectroscopy (EELS) analysis. EELS elemental mapping was obtained using the 'three window technique'.[15] Figures 4(a) and 4(b) show EELS energy-filtered oxygen maps and corresponding intensity profiles of the regions marked with white rectangular areas, visualizing the distribution of oxygen in the off and the on states specimens, respectively. One remarkable finding is the oxygen deficiency existing in the middle region of titanium oxides. Many researchers have believed that oxygen vacancies at the metal-oxide interface perform the critical function of bistable resistive switching. However, from the above the EELS result, it is found that the oxygen concentration at the interface very close to aluminum top electrode is much higher than that of the inner part, as displayed in red in the color scale. This can be explained via an oxygen diffusion process. When the interaction between aluminum and titanium oxide occurred at the top interface, the insufficient oxygen ions was added continuously from the below bulk domain. As a result, the bulk region underneath the top interface became more oxygen deficient than the top interface. Thus, it is suggested that these oxygen vacancies present on the top bulk domain become a huge factor in the charge trapping sites and cause bulk space-charge-limited-conduction (SCLC). In particular, the EELS oxygen mapping image of the on state, shown in Figure 3b and 3d, remarkably shows that oxygen ions driven into the top interface region are equally distributed through the titanium oxide inner part. It can be believed that oxygen ions are easily able to move in the $TiO_{2-x}$ amorphous active layer by strong external electric field.

Based on above results, the bipolar switching mechanisms of our $Al/TiO_2/Al$ devices can be described schematically in Figure 4(c). When the top Al electrode deposition started in the thermal evaporation chamber and aluminum atoms attached to the surface of the titanium oxide, they gathered oxygen ions present on the top $TiO_2$ layer, resulting in the formation of the top interface amorphous layer owing to strong oxygen affinity of Al metal. The top interface may be Al-Ti-O phases. Therefore, the oxygen deficient domain was created at the bulk $TiO_2$ region. This state corresponds to the off state in $Al/TiO_2/Al$ memory devices. While sufficient negative bias is applied to the top Al electrode, the negatively charged oxygen ions piled up at the vicinity of Al-Ti-O layer diffuse into the inner bulk $TiO_{2-x}$ region, causing the device to switch to the on state.

In summary, we have demonstrated that the drift of oxygen ions (vacancies) by the applied bias was attributed to the microscopic origin of the bistable resistivity switching of Al/$TiO_2$/Al devices. From the sequential cross-sectional HRTEM images, *in-situ* XPS measurement, and energy-filtered TEM elemental mapping analysis, it is justified that oxygen ions piled up at top interface region can drift as external bias exceeds the threshold field. These analyses give the direct evidence of the memristic behaviors and the deeper understanding of the bipolar resistive switching mechanisms of binary oxides.

**Acknowledgments**

This work was supported by the Next-generation Non-volatile Memory Program of the Ministry of Knowledge Economy, Korea. The authors sincerely thank Youn-Joong Kim at Korea Basic Science Institute (KBSI) for the use of the high-voltage electron microscope and the technical staffs of the Process Development Team at ETRI for their support with the PEALD facility.

**References**

[1]J. F. Gibbons and W. E. Beadle, Solid-St. Electron. **7,** 785 (1964).

[2]W. R. Hiatt and T. W. Hickmott, Appl. Phy. Lett. **6**, 106 (1965).

[3]F. Argall, Electron. Lett. **2**, 282 (1966).

[4]R. W. Brander, D. R. Lamb, and P. C. Rundle, Brit. J. Appl. Phys. **18**, 23 (1967).

[5]F. Argall, Solid-St. Electron. **11**, 535 (1968).

[6]I. G. Baek, M. S. Lee, S. Seo, M. J. Lee, D. H. Seo, D.-S. Suh, J. C. Park, S. O. Park, H. S. Kim, I. K. Yoo, U-In Chung, and J. T. Moon, Dig. − Int. Electron Devices Meet. **2004**, 587.

[7]I. G. Baek, D. C. Kim, M. J. Lee, H.-J. Kim, E. K. Yim, M. S. Lee, J. E. Lee, S. E. Ahn, S. Seo, J. H. Lee, J. C. Park, Y. K. Cha, S. O. Park, H. S. Kim, I. K. Yoo, U-In Chung, J. T. Moon, and B. I. Ryu, Dig. − Int. Electron Devices Meet. **2005**, 750.

[8]M.-J. Lee, S. Seo, D.-C. Kim, S.-E. Ahn, D. H. Seo, I.-K. Yoo, I. G. Baek, D.-S. Kim, I.-S. Byun, S.-H. Kim, I.-R. Hwang, J.-S. Kim, S.-H. Jeon, and B. H. Park, Adv. Mater. **19**, 73 (2007).

[9]M.-J. Lee, Y. Park, D.-S. Suh, E.-H. Lee, S. Seo, D.-C. Kim, R. Jung, B.-S. Kang, S.-E. Ahn, C. B. Lee, D. H. Seo, Y.-K. Cha, I.-K. Yoo, J.-S. Kim, and B. H. Park, Adv. Mater. **19**, 3919 (2007).

[10]J. J. Yang, M. D. Pickett, X. Li, D. A. A. Ohlberg, D. R. Stewart, and R. S. Williams, Nature Nanotech. **3**, 429 (2008).

[11]D. B. Strukov, G. S. Snider, D. R. Stewart, and R. S. Williams, Nature **453**, 80 (2008).

[12]L.-E. Yu, S. Kim, M.-K. Ryu, S.-Y. Choi, and Y.-K. Choi, IEEE Electron Device Lett. **29**,

331 (2008).

[13]U. Diebold, Surf. Sci. Rep. **48**, 53 (2003).

[14]L. S. Dake, R. J. Lad, Surf. Sci. **289**, 297 (1993).

[15]L. Reimer, Energy-Filtering Transmission Electron Microscopy, Springer Verlag, Berlin, 387 (1995).

**Figure Captions**

FIG. 1. Current density-voltage (J-V) characteristics of a typical Al/TiO$_2$/Al memory device. Cross-sectional BF TEM image of Al/TiO$_2$/Al sandwiched structure deposited on a SiO$_2$ (3000 Å)/Si substrate is shown in the right bottom inset.

FIG. 2. (Color online) A series of the cross-sectional HRTEM images of Al/TiO$_2$/Al heterostructures observed step by step: (a) Image showing the existence of native aluminum oxide at the aluminum metal surface. (b) Image indicating the broadening of native aluminum oxide by the thermal annealing. (c) Image obtained after depositing amorphous titanium oxide thin film by a 400 cycle PEALD process. (d) Completed cell image acquired by thermal evaporation deposition of the aluminum top electrode.

FIG. 3. (Color online) XPS spectra of (a) Ti 2p and (b) Al 2p measured as a function of Al thickness on amorphous titanium oxide film deposited by the PEALD process.

FiG. 4. (Color online) Energy filtered-TEM elemental (oxygen) maps of two different samples (the off and on resistance state): (a) EELS based oxygen elemental map of as-grown Al/TiO$_2$/Al structure, corresponding to the off state. (b) Oxygen elemental map of the sample applied negative set bias (-3 V) on the top Al electrode, corresponding to the on state. The left insets are the corresponding average intensity profiles obtained from the white rectangular areas. (c) Schematics of proposed models for bipolar resistive switching of Al/TiO$_2$/Al device.

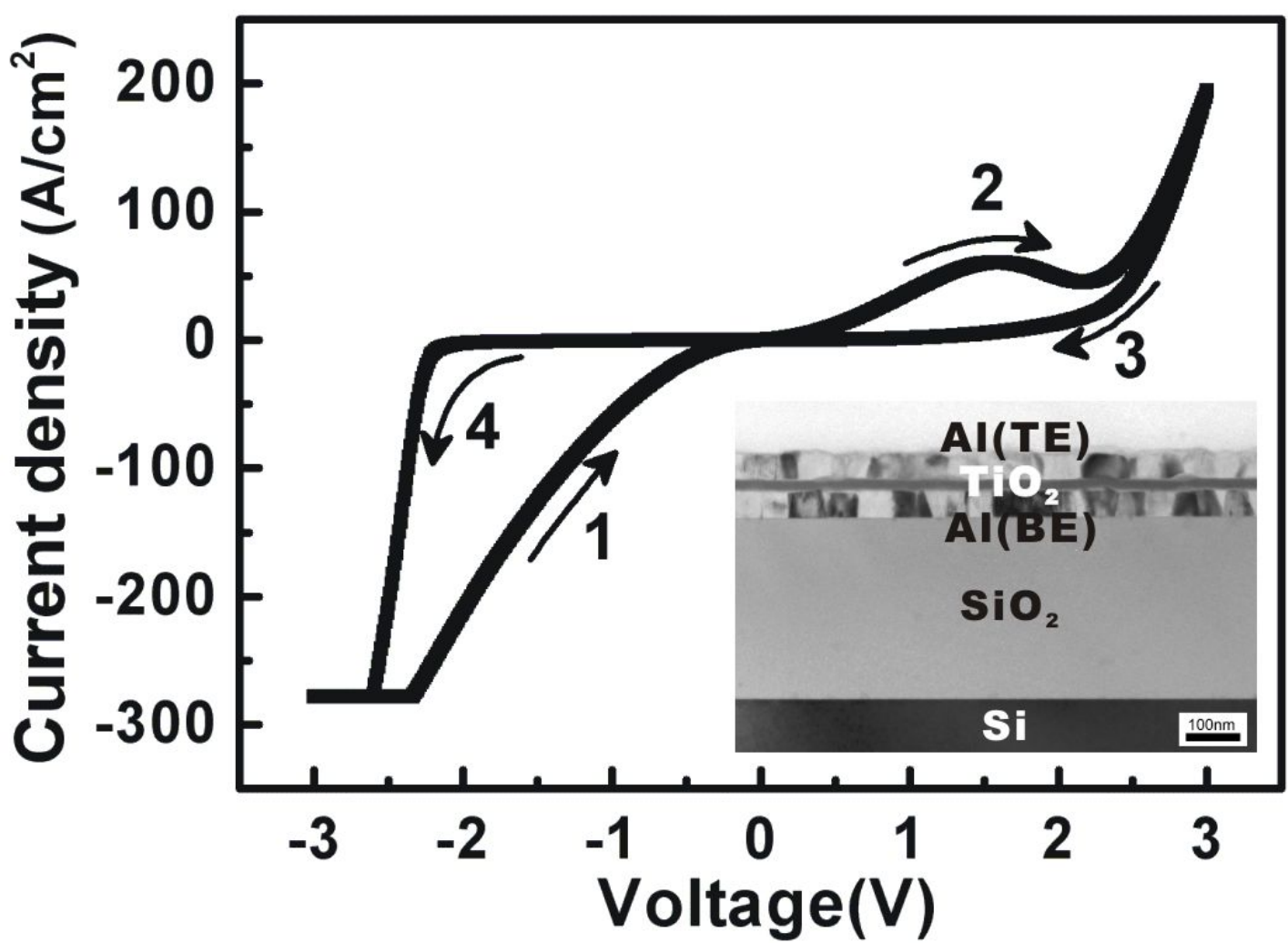

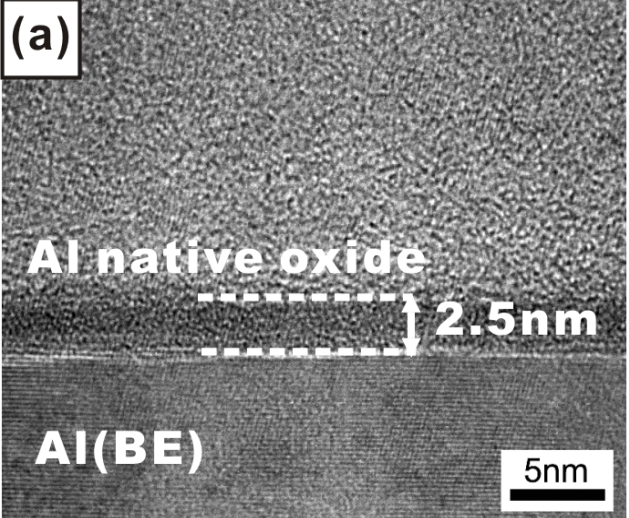
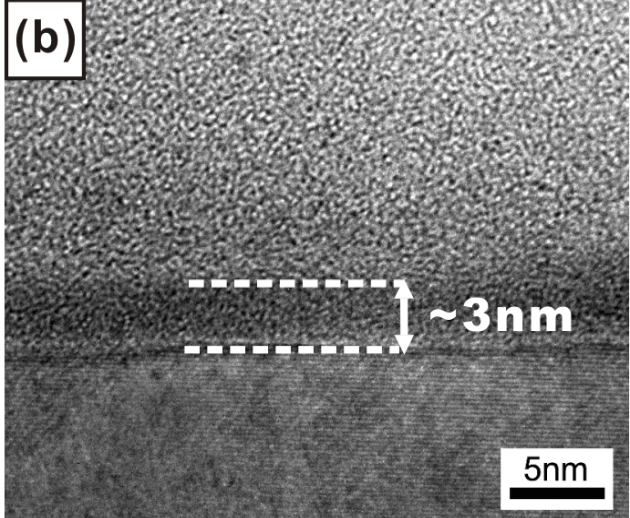
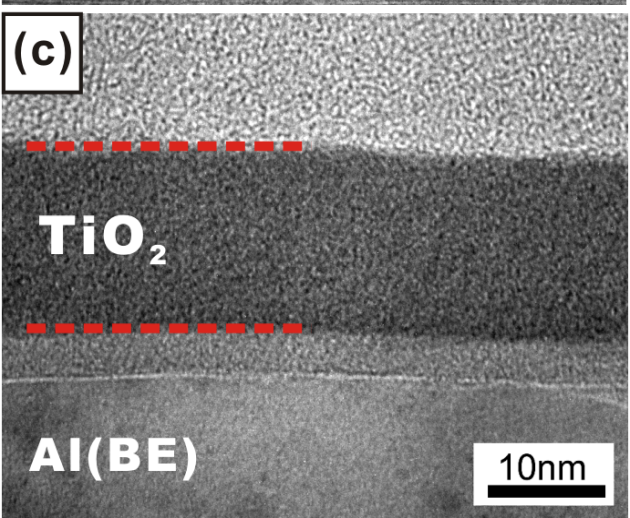
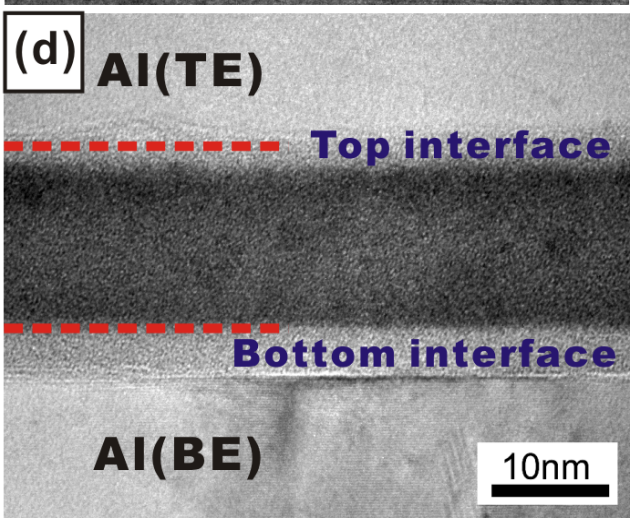

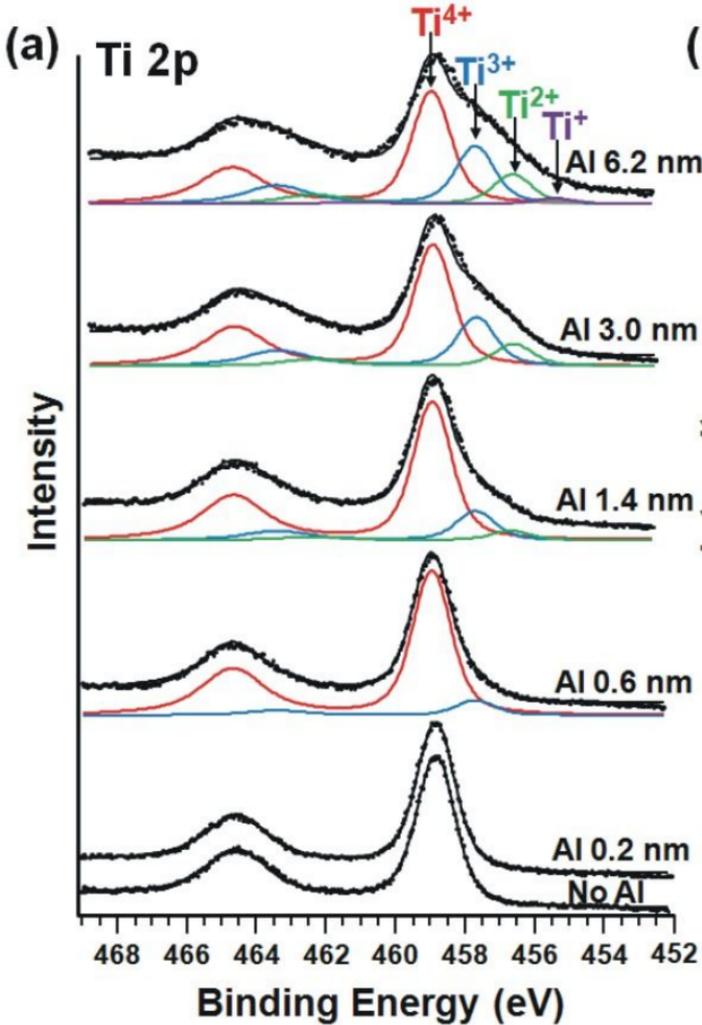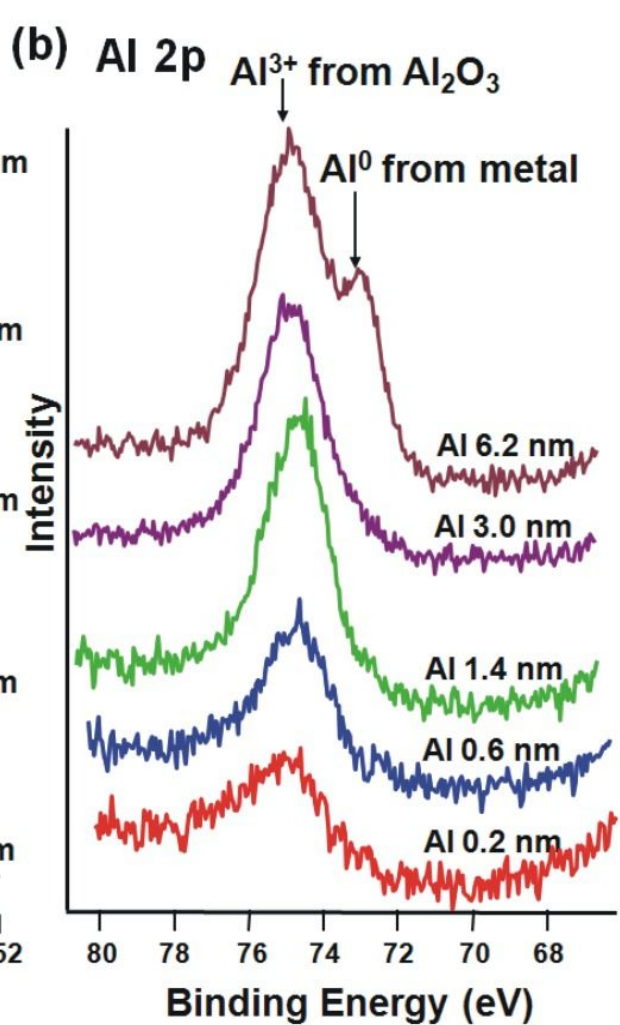

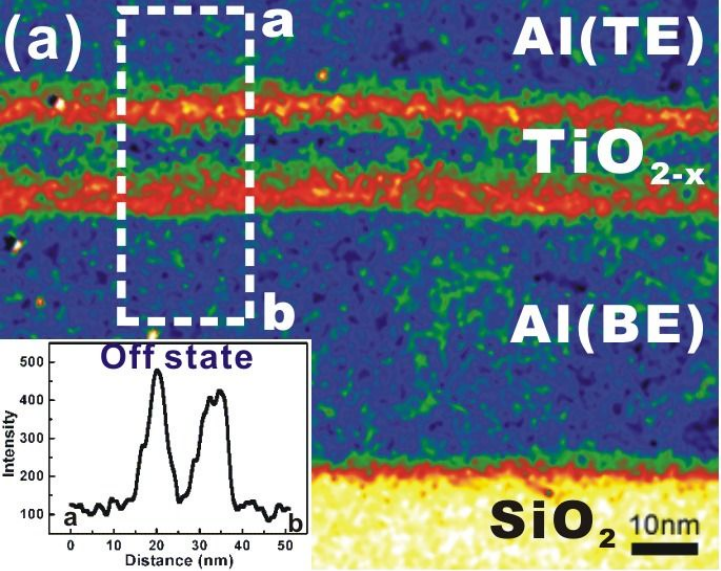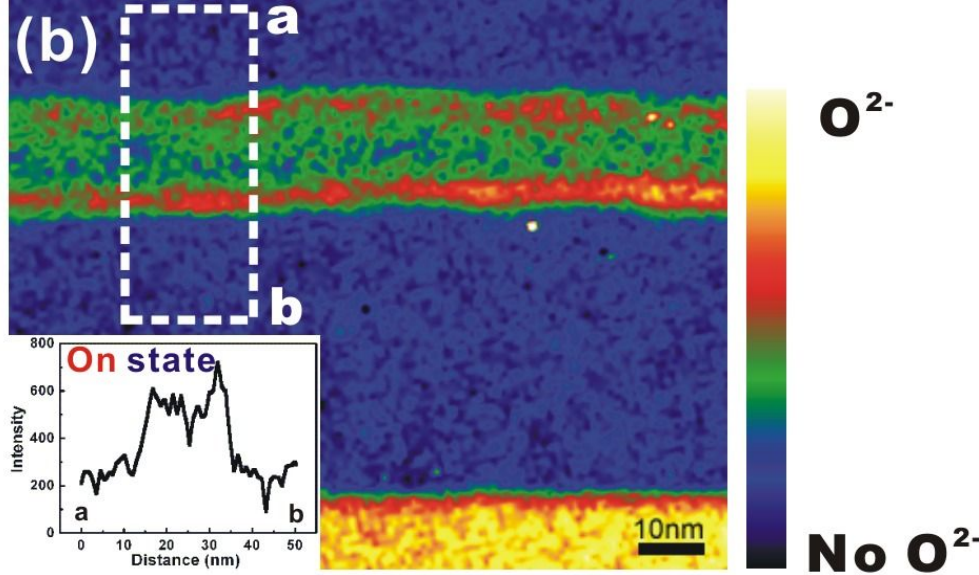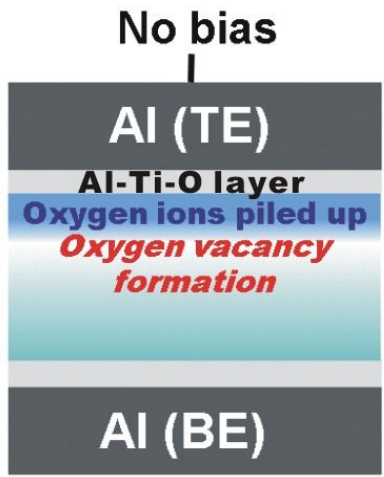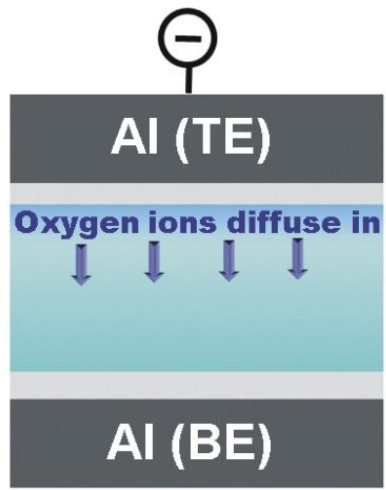